\documentstyle[prd,aps,twocolumn,epsf,axodraw]{revtex}
\def\br{\begin{eqnarray}}
\def\er{\end{eqnarray}}
\def\be{\begin{equation}}
\def\ee{\end{equation}}

\def\L{\Lambda}

\def\({\left(}
\def\){\right)}

\begin{document}
\twocolumn[\hsize\textwidth\columnwidth\hsize\csname 
@twocolumnfalse\endcsname                            

%
%
\title{Freezing of the QCD coupling constant and solutions
of Schwinger-Dyson equations}
\author{A.~C.~Aguilar, A.~Mihara and A.~A.~Natale\\}
\address{Instituto de F\'{\i}sica Te\'orica\\
Universidade Estadual Paulista\\
Rua Pamplona 145\\
01405-900, S\~ao Paulo, SP\\
Brazil}
%
\date{\today}
\maketitle

\begin{abstract}

We compare phenomenological values of the frozen QCD running
coupling constant ($\alpha_s$) with two classes of solutions
obtained through nonperturbative Schwinger-Dyson equations. We
use these same solutions with frozen coupling constants as well
as their respective nonperturbative gluon propagators to compute
the QCD prediction for the asymptotic pion form factor. Agreement
between theory and experiment on $\alpha_s(0)$ and $F_\pi (Q^2)$
is found only for one of the solutions Schwinger-Dyson equations.

\end{abstract}


\vskip 0.5cm]                            

\section{Introduction}

\noindent

The possibility that the QCD coupling constant ($\alpha_s$) has an
infrared (IR) finite behavior has been extensively studied in
recent years. There are theoretical arguments in favor of the
coupling constant freezing at low momenta, one of them, \`a la
Banks and Zaks\cite{bz}, claims that QCD may have a non-trivial
IR fixed point even for a small number of quarks (see, for
instance, Ref.\cite{grunberg}). We can also use arguments of
analyticity to show that the analytical coupling freezes at the
value of $4\pi/ \beta_0$\cite{shirkov}, where $\beta_0$ is the
one-loop coefficient of the QCD $\beta$ function.

Studies of the nonperturbative QCD vacuum also indicate the
existence of a finite coupling constant in the
IR\cite{simonov,gribov}.

The phenomenological evidences for the strong coupling constant
freezing in the IR are much more numerous. Models where a static
potential is used to compute the hadronic spectra make use of a
frozen coupling constant at long distances\cite{eichten,isgur}.

Heavy quarkonia decays and total hadron-hadron cross sections are
influenced by the freezing of the coupling constant\cite{parisi}.

A quite detailed analysis of the ratio $R_{e^+e^-}$ ($\equiv
\sigma_{tot} (e^+ e^- \rightarrow hadrons)/\sigma ( e^+ e^-
\rightarrow \mu^+ \mu^-)$)performed by Mattingly and
Stevenson\cite{stevenson} also shows a signal for the freezing of
the QCD coupling. Following an almost similar study, for several
hadronic observables, Dokshitzer and Webber obtain the same
result\cite{dokshitzer}.

Another method to investigate the infrared behavior of gluon and
ghost propagators, and of the running coupling constant at low
energies is through the solution of the Schwinger-Dyson equations
(SDE)\cite{roberts}. Early studies of the SDE for the gluon
propagator in the Landau gauge concluded that the gluon
propagator is highly singular in the infrared\cite{mandelstam}.
However, these results are discarded by simulations of QCD on the
lattice at $95 \%$ confidence level\cite{lat}, where it is shown
that the gluon propagator probably is infrared finite. The
lattice result is in agreement with two classes of SDE solutions.
One proposed by Cornwall many years ago where the gluon acquires
a dynamical mass\cite{cornwall}, and another that has been
extensively discussed by Alkofer and von Smekal where the gluon
propagator goes to zero when the momentum $q^2 \rightarrow
0$\cite{alkofer}. The solutions differ due to the different
approximations performed to solve the SDE, but in both cases
there is a freezing of the coupling constant in the IR.

Although the figures of the most recent lattice
calculation\cite{lat} seems to indicate that the Cornwall's gluon
propagator is the one that could better explain the results, it
is correct to say that the data is still not precise enough in
the IR region to decide among the two possible behaviors for the
gluon propagators discussed in the previous paragraph. The purpose
of our work is exactly to confront the IR values of the
theoretical coupling constant, obtained with the solutions of the
SDE, with the phenomenological data about the value of $\alpha_s
(0)$ in order to discriminate which one is the most suitable
solution. Finally, these theoretical and phenomenological
calculations are outside the scope of standard perturbation
theory, and a consistency check between them is the minimum that
we may require to know if these approaches make sense at all. In
the next section we present the expressions of the
nonperturbative running coupling constant obtained with the SDE
study, and compare them with some of the phenomenological values
obtained for $\alpha_s (0)$. In Section III we compute the pion
form factor ($F_\pi (Q^2)$) as a function of these coupling
constants. It is known that $F_\pi (Q^2)$ is quite dependent on
the behavior of $\alpha_s $ at small momentum\cite{ji}.

Therefore, this calculation provides a good test for the
nonperturbative expression of the QCD coupling constant. Considering
that solutions of SDE show a nonperturbative behavior for
the infrared coupling as well as for the gluon propagators,
in Section IV we modify the expression for the asymptotic
pion form factor to take into account these nonperturbative
gluon propagators. In the last section we present our
discussion and conclusions.

\section{The phenomenological value of $\alpha_s (0)$}

At high energies it is believed that the property of asymptotic
freedom allow us to perform reliable QCD calculations. However,
the same is not true at low energies, where we have to make use
of a series of phenomenological models when computing strong
interaction parameters.

This is exactly what happens if we want to determine the IR
behavior of the running coupling constant. We are going to
present some of the determinations of $\alpha_s (0)$, and the
most impressive fact is that the values obtained in several
different analysis are not far apart by one order of magnitude,
but they differ at most by a factor of two, providing a solid
indication of the robustness of these approaches.

One of the most detailed calculation of $\alpha_s (0)$ is due to
Mattingly and Stevenson\cite{stevenson}, which uses perturbation
theory and renormalization group invariance to compute
$R_{e^+e^-}$ up to third order in $\alpha_s$. They predict the
value
\begin {equation}
\alpha_s / \pi = 0.26,
\label{ms}
\end{equation}
($\alpha_s (0) = 0.82$) for the frozen IR coupling. On the other
hand the long work of Ref.\cite{dokshitzer} gives
\begin {equation}
\alpha_s (0) \approx 0.63.
\label{do}
\end{equation}
The analysis of hadronic spectroscopy with potential models by
Godfrey and Isgur\cite{isgur} led to the following behavior of
the coupling constant
\begin {eqnarray}
\alpha_{gi} = 0.25 \exp (&&-q^2) + 0.15 \exp (-0.1q^2) \nonumber\\
&&+ 0.20 \exp (-0.001q^2), \label{gi}
\end{eqnarray}
where $q$ is in GeV (all the momenta, otherwise specified, will be
in Euclidean space), and a good fit of the spectra does not
depend strongly on the ultraviolet behavior of the coupling
constant. From the above equation we obtain $\alpha_s (0) =
0.60$. Which is also consistent with more recent studies of QCD
potentials\cite{jezabek}. Analysis of $e^+e^-$ annihilation, as
well as bottomonium and charmonium fine structure in the
framework of the background perturbation theory may lead to a
frozen value of the coupling constant as low as $\alpha_s (0)
\approx 0.4$\cite{badalian}. This method also explains the frozen
value of $\alpha_s$ resulting from the lattice simulation of the
short range static potential\cite{badalian2}, and it gives
\begin {eqnarray}
\alpha_{B}(0) \approx \frac{4\pi}{\beta_0
\ln{\frac{m^2_B}{\Lambda^2_V}}},
\label{gb}
\end{eqnarray}
where $m_B$ is a background mass. This one and $\Lambda_V$
(with $m_B>\Lambda_V$) are determined
phenomenologically\cite{badalian2}.

There are many other results that we could present here, but we
can assume that the phenomenological values of $\alpha_s (0)$
scattered in the literature are in the range
\begin {equation}
\alpha_s (0) \approx 0.7 \pm 0.3.
\label{aver}
\end{equation}
Although this choice is {\it ad hoc}, as far as we know it
contemplates most of the phenomenological determinations of
$\alpha_s (0)$.

We now turn to the coupling constants obtained through the SDE
solutions. The first nonperturbative running coupling constant
that we shall discuss was obtained by Cornwall\cite{cornwall},
using the pinch technique to derive a gauge invariant SDE. This
nonperturbative coupling is equal to
\be \alpha_{sC} (q^2)= \frac{4\pi}{\beta_0 \ln\left[
(q^2+4M_g^2(q^2)/\L^2 \right]}, \label{acor} \ee
where $M_g(q^2)$ is a dynamical gluon mass given by,
\be M^2_g(q^2) =m_g^2 \left[\frac{ \ln \(\frac{q^2+4{m_g}^2}{\Lambda
^2}\) } { \ln\(\frac{4{m_g}^2}{\Lambda ^2}\) }\right]^{- 12/11}
\label{mdyna} \ee
$\L$($\equiv\L_{QCD}$) is the QCD scale parameter, $\beta_0 = 11
- \frac{2}{3}n_f$, where $n_f$ is the number of flavors. In the
above expression we are neglecting the effect of dynamical or
bare fermions masses\cite{cornwall}. We can determine $\alpha_s
(0)$ in Eq.(\ref{acor}) as a function of the gluon mass $m_g$ and
$\L$, and these ones can be obtained in the calculation of
several hadronic parameters that may vary with $m_g$ (but, in
general, not strongly with the ratio $m_g / \L$). A typical value
is\cite{cornwall,natale}
\begin {equation}
m_g = 500 \pm 200 \quad {\textnormal MeV}
\label{mg}
\end{equation}
for $\L = 300$ MeV. It is interesting to observe the similarity
between Eq.(\ref{acor}) and Eq.(\ref{gb}). Although, it is not
clear to us the reason for this similarity.

The other possibility for the IR finite running coupling
was studied by Alkofer et al. \cite{alkofer}, that solved a
coupled set of SDE for the propagators of gluons and ghosts.
In this approach the solution to the running coupling leads
to an infrared fixed point, which, in terms of the invariant
functions $Z(k^2)$ and $G(k^2)$ related to the
renormalization of gluon and ghost propagators respectively,
is given by
\br \alpha_s (\mu)= \frac{g^2}{4\pi \beta_0} Z(\mu^2)G^2 (\mu^2)
|_{\mu \rightarrow 0} \nonumber \\
= \frac{16\pi}{3N_c} \left( \frac{1}{\kappa} - \frac{1}{2}
\right)^{-1} \simeq 9.5. \label{alkoir}
\er
with ${\kappa}= 0.92 $.

The above result gives $\alpha_s$ near the origin. It has been
obtained for $n_f=0$. As we are going to compare different SDE
solutions we will limit ourselves to the flavorless, or pure
gauge, QCD. The effect of $n_f \neq 0$ will be discussed in the
last section.

Since we shall use the running coupling in the full range of
momenta we provide a fit for the numerical data of
Ref.\cite{alkofer}, given by the following expression:
\br
\alpha_{sA}=\left\{\begin{array}
{l@{\quad:\quad}l}
\alpha_{sA1}&   q^2 < 0.31 \;  {\textnormal GeV}^2\\
\alpha_{sA2}& 0.31 < q^2 < 1.3 \; {\textnormal GeV}^2\\
\alpha_{sA3} & q^2>1.3\;  {\textnormal GeV}^2
\end{array} \right.
\label{ralk}
\er
with

\br
\alpha_{sA1}&=&   0.2161 + 9.2621
\exp\(-2\frac{(q^2-0.0297)^2}{(0.6846)^2}\)      \nonumber
\\ \alpha_{sA2}&=&  1.4741 + 8.6072
\exp\(-\frac{q^2-0.1626}{0.3197}\)      \nonumber   \\
\alpha_{sA3}&=&  \frac{1.4978}{\ln (1.8488 q^2)} , \er %
where the $\chi^2 \approx  2.5\times 10^{-4}$  for the three regions.


\begin{figure}
\setlength{\epsfxsize}{1.0\hsize} \centerline{\epsfbox{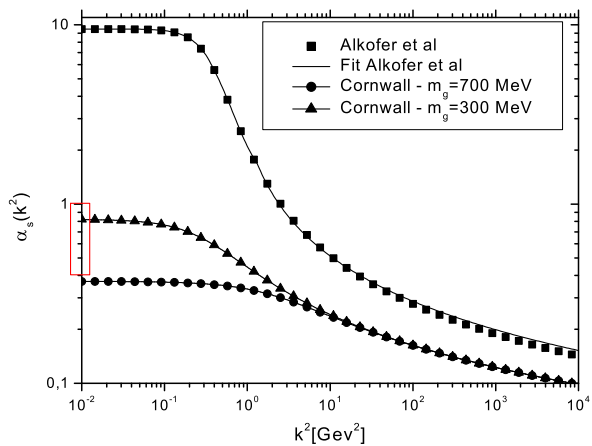}}
\caption[dummy0] {Comparison between the running couplings
obtained from different approximations in the SDE study. The
curves with line + triangle and line + circle  delimit  the
phenomenological range acceptable for the gluon mass ($300$\, MeV
and $700$\, MeV, respectively) with  $\Lambda = 300$\,MeV for the
Cornwall's running coupling. The square points are the numerical
data computed by Alkofer et al. and the solid line is our fit
Eq.(\ref{ralk}). The box in y axis shows the phenomenological
range indicated in Eq.(\ref{aver}). We can see that only the
running coupling computed by Cornwall is  compatible with the
phenomenological estimatives of $\alpha_s (0)$.} \label{running}
\end{figure}

%

In Fig.(\ref{running}) we indicate the expected phenomenological
range of values for $\alpha_s (0)$ and plot the curves for
$\alpha_{sC}$ and $\alpha_{sA}$.

It is evident that only the Cornwall's solution is
compatible with the phenomenological data. In the last
section we shall comment on possible modifications of this
result.

\section{The nonperturbative coupling and the pion form factor}

It is known that the pion form factor, $F_\pi (Q^2)$, is quite
dependent on the behavior of $\alpha_s $ at small
momentum\cite{ji}. The asymptotic form factor is predicted by
perturbative QCD\cite{ji,brodsky}. It depends on the internal pion
dynamics that is parametrized by the quark distribution amplitude
of the pion. The QCD expression for the pion form factor
is\cite{brodsky}
\br F_{\pi}(Q^2)= \int_{0}^{1}\!\!dx \! \int_{0}^{1}\!\!dy \,&&
\phi^{*}
(y,\tilde{Q}_y) T_H(x,y,Q^2) \nonumber\\
&&\times\phi(x,\tilde{Q}_x),
 \label{fpi} \er
where $\tilde{Q}_x= \min(x,1-x)Q$ and Q is the 4-momentum in Euclidean space transferred by the photon . The
function $\phi(x,\tilde{Q}_x)$ is the pion wave function,
that gives the amplitude for finding the quark or antiquark
within the pion carrying the fractional momentum $x$ or
$1-x$, respectively. In this work we use the model for the
pion distribution amplitude proposed by Chernyak and
Zhitnitsky \cite{Chernyak}. This wave function was derived
from QCD sum rules and it is written as
\br \phi(x,Q) = &&\frac{f_{\pi}}{2\sqrt{3}}x(1-x)\nonumber \\
&&\times \left\{6+\left[ 30(2x-1)^2-6
\right]\left(\frac{\alpha_s(Q)}{\alpha_s(\mu)}\right)^{{\gamma}
_2} \right\}, \label{phi}
\er
with $\mu= 500 \,$ MeV and
\be
\gamma_{2} =\frac{50}{99-6n_f}.
\label{gamma2}
\ee

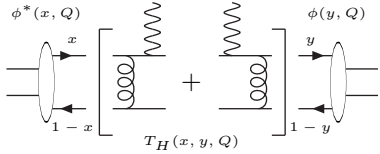
\begin{figure}[ht]
\begin{center}\begin{picture}(160,90)(0,0)
\Line(10,55)(25,55) \Line(10,45)(25,45) \ArrowLine(25,60)(40,60)
\ArrowLine(40,40)(25,40) \GOval(25,50)(15,3)(0){1}
\Text(35,65)[cb]{{\tiny $x$}} \Text(35,35)[ct]{{\tiny $1-x$}}
\Text(25,75)[cb]{{\tiny $\phi^{*}(x,Q)$}}
%
\Line(45,70)(50,70) \Line(45,30)(50,30) \Line(45,30)(45,70)

\Line(50,60)(70,60) \Line(50,40)(70,40)
\Photon(65,80)(65,60){3}{4} \Gluon(55,60)(55,40){3}{3}
\Text(80,50)[]{+} \Line(90,60)(110,60) \Line(90,40)(110,40)
\Photon(95,80)(95,60){3}{4} \Gluon(105,60)(105,40){3}{3}

\Text(80,30)[ct]{{\tiny $T_{H}(x,y,Q)$}} \Line(110,70)(115,70)
\Line(110,30)(115,30) \Line(115,30)(115,70)
%

\Line(135,55)(150,55) \Line(135,45)(150,45)
\ArrowLine(120,60)(135,60) \ArrowLine(135,40)(120,40)
\GOval(135,50)(15,3)(0){1} \Text(125,65)[cb]{{\tiny $y$}}
\Text(125,35)[ct]{{\tiny $1-y$}} \Text(135,75)[cb]{{\tiny
$\phi(y,Q)$}}
\end{picture}\end{center}
\caption[dummy0]{The leading-order diagrams that contribute to
the pion form factor. $\phi(x,\tilde{Q}_x)$ is the pion wave
function, that gives the amplitude for finding the quark or
antiquark within the pion carrying the fractional momentum $x$ or
$1-x$. The photon transfers  the momentum $q^\prime$ (in Minkowski
space), $Q^2=-q^{\prime 2}$,  for the  $q\overline{q}$ pair of
total momentum $P$ producing a $q\overline{q}$ pair of final
momentum $P^{\prime}$.} \label{PFF}
\end{figure}

 The other function, $ T_H (x,y,Q^2)$, is the hard-scattering
amplitude that is obtained by computing the quark-photon
scattering diagram as shown in Fig. {\ref{PFF}}. The lowest-order
expression of $ T_H(x,y,Q^2)$ is given by (see \cite{braaten},
and references therein)
\br T_{H}(x,y,Q) &=& \frac{64\pi}{3Q^2}\left\{\frac{2}{3}
\frac{\alpha_s[(1-x)(1-y)Q^2]}{(1-x)(1-y)} \right. \nonumber\\
&& + \left.\frac{1}{3} \frac{\alpha_s(xyQ^2)}{xy} \right\}
\label{thard} \er
To compute the pion form factor using the nonperturbative
runningbcouplings proposed by Alkofer {\it et
al}.\cite{alkofer} and Cornwall Eq.(\ref{acor}), we solved
the integrals given by Eq.(\ref{fpi}) substituting the quark
distribution amplitude written in Eq.(\ref{phi}) and the
expression of $ T_H(x,y,Q^2)$ (Eq.(\ref{thard})).

\begin{figure}
\setlength{\epsfxsize}{1.0\hsize} \centerline{\epsfbox{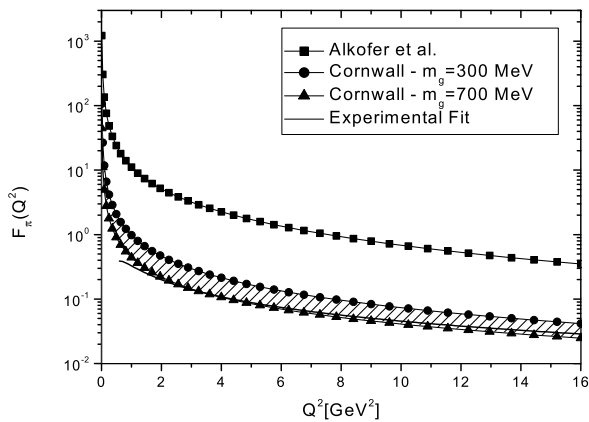}}
\caption[dummy0]{Pion form factor computed with the different
nonperturbative running coupling constants. The curve composed by
line + square is obtained with  Eq.(\ref{ralk}). The curves that
define the shaded area are computed with Eq.(\ref{acor}) for the
values of the $m_g=300$\, (upper curve) and $700$\,MeV (lower
curve). The solid line is the experimental data fit
Eq.(\ref{fpiex}). There is a nice agreement when $F_{\pi}$ is
computed with Cornwall's running coupling.} \label{fpis}
\end{figure}

The pion form factor result for the different forms of the QCD
coupling in the low momentum regime is shown in Fig.(\ref{fpis}).
We used for Eq.(\ref{acor}), the lower\,($300 \,$ MeV) and the
upper\,($700\,$ MeV) gluon mass values for a fixed
$\Lambda=300\,$ MeV. These values defined the shaded area
representing the expected range for the pion form factor,
$F_{\pi}$.

%


In this same figure, we also compare our results with the
experimental data (solid line) \cite{exp} that was described by
the least $\chi^2$ fit ($\chi^2_{min}=7.96742$) determined in
Ref.\cite{yeh}
\be F^{fit}_{\pi} = \frac{0.46895}{Q^2}\left( 1 -
\frac{0.3009}{Q^2} \right). \label{fpiex}
\ee
The results, using the running coupling of the Eq.(\ref{acor}),
agree very nicely with the experimental data for a gluon mass
value close to $700\,$MeV. On the other hand, the calculations
with Eq.(\ref{ralk}) overestimate $F_{\pi}$ at least by one order
of magnitude.


\section{EFFECTS OF NONPERTURBATIVE PROPAGATORS IN THE $F_{\pi}$ BEHAVIOR}

In the previous section we computed $F_{\pi}$ using two distinct
forms of the  nonperturbative running coupling. We considered
that the gluon exchanged by the $q\overline{q}$ pair of
Fig.(\ref{PFF}) is a perturbative one. However, the SDE solutions
at the same time that they give the nonperturbative behavior of
the running coupling, they provide nonperturbative expressions
for the gluon propagators that at the origin differ drastically
from the perturbative propagator.

The large momentum behavior of these nonperturbative propagators
coincide with the perturbative one, and, by consistency,
we have to use the nonperturbative gluon propagators together with
their respective coupling constants, even considering that
we are computing the asymptotic pion form factor. So that,
it is worth asking whether our previous analysis would be
distinct if we change the perturbative gluon propagator by
the full one.

In order to introduce this modification, we verify that in
Eq.(\ref{thard}) we used the perturbative QCD gluon propagator
that, in the Landau gauge, is given by
\be D_{\mu\nu}(q^2)= \({\delta}_{\mu\nu}
-\frac{q_{\mu}q_{\nu}}{q^2}\)D(q^2), \quad D(q^2)=\frac{1}{q^2}.
\label{landau} \ee
We can easily factorize $D(q^2)$ in the Eq.(\ref{thard}),
rewritten this last equation as
\be T_H(x,y,Q^2) = \frac{64\pi}{3}\left[
\frac{2}{3}\alpha_s(K^2)D(K^2) +
\frac{1}{3}\alpha_s(P^2)D(P^2)\right], \label{thD} \ee
where $K^2=(1-x)(1-y)Q^2$ and $P^2=xyQ^2$.

Let us now consider the two different nonperturbative behaviors
of gluon propagators. The first one was obtained by
Cornwall\cite{cornwall}, and is given by
\be
D_C(q^2)=\frac{1}{q^2 + M_{g}^2(q^2)}.
\label{propcorn}
\ee
where $M_g(q^2)$ is the dynamical mass given by
Eq.(\ref{mdyna}).  The gluon propagador computed by Alkofer
{\it et  al.}\cite{alkofer}, can be fitted by the following
expression  (${\chi}^2=0.016$)
\be D_A(q^2)=\frac{bq^2}{q^4 + a^4}, \label{propalk}
\ee
where $a=0.603$ and $b=3.707$.

 Once the propagators are given by Eqs.(\ref{propcorn},\ref{propalk})then $T_H$
(Eq.(\ref{thD})) will be changed to
\br T_H(x,y,Q^2) &=& \frac{64\pi}{3}\left[
\frac{2}{3}\alpha_s(K^2)D_{A,C}(K^2) \right.\nonumber\\
& & \left. + \frac{1}{3}\alpha_s(P^2)D_{A,C}(P^2)\right].
\label{thDAC}
\er

\begin{figure}[ht]
\setlength{\epsfxsize}{1.0\hsize} \centerline{\epsfbox{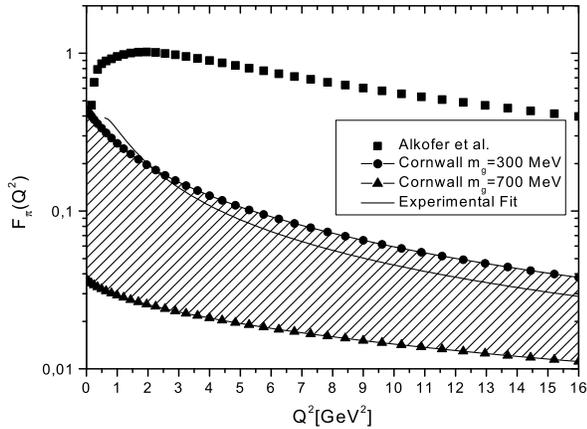}}
\caption[dummy0]{Comparison among the experimental curve (solid
line) and our results for the pion form factor using the
Alkofer's and Cornwall's nonperturbative propagators and coupling
constants.} \label{fpi-dif}
\end{figure}

We performed the integrations of Eq.(\ref{fpi})
numerically, with the amplitude $T_H$ given by
Eq.(\ref{thDAC}) and their respective running coupling
constant (see Fig.(\ref{running})).
Our results are shown in Fig.(\ref{fpi-dif}).


If we compare the results of Fig.(\ref{fpi-dif}) with the results
of the previous section, we can observe a striking attenuation of
$F_{\pi}(Q^2)$ for $Q^2 \rightarrow 0$. It is also clear that
$F_{\pi}(0)$ is finite for both models. This new behavior at low
momentum can be understood if we notice that, for $Q^2
\rightarrow 0$,
\br
D_A(Q^2) &\rightarrow& 0 \\
D_C(Q^2) &\rightarrow& \mbox{finite},
\er
in contrast with the divergent perturbative propagator. When we
use the nonperturbative information obtained through the SDE with
the approximations of Ref.\cite{alkofer}, we continue to have a
disagreement with the experimental data. In this particular case
the pion form factor even go to zero as $Q^2 \rightarrow 0$.

Obviously we should not consider this region of tranferred
momentum because the kernel of Eq.(\ref{thDAC}) is valid only for
large $Q^2$, but the disagreement goes through the asymptotic
region. On the other hand, the Cornwall's propagator is still
compatible with the experimental data, but now the agreement is
in favor of smaller gluon masses.

\section{Discussion and conclusions}

There is an increasing phenomenological evidence for the freezing
of the QCD running coupling constant in the infrared region. It
is clear that much more work has to be done in order to establish
definitively these results. However, it is very satisfying to see
that they are not far apart, and are concentrated on a region
slightly below $\alpha_s \approx 1$.

On the theoretical side there are many studies leading also to
this infrared fixed point. Among these we selected the ones
derived from the solutions of Schwinger-Dyson equations.

In this work we proposed to test the compatibility between the
phenomenological values of $\alpha_s (0)$ with the values given
by the SDE solutions. This compatibility (or not) can teach us if
the approximations used to solve the SDE are realistic or not,
and if more data is accumulated we may even be able to discard
nonphysical solutions.

We discussed two SDE solutions for the running coupling constant
and gluon propagators. One proposed in Ref.\cite{cornwall} and
the other in Ref.\cite{alkofer}. These are the only ones
consistent with the recent simulations on the lattice of the
gluon propagator\cite{lat}. These solutions have been obtained in
Euclidean space and in pure gauge QCD, i.e. $n_f=0$.

The effect of the number of flavors in Cornwall's
solution\cite{cornwall} is not so strong, and it appears in the
coefficient $\beta_0$ of the coupling constant and in the gluon
mass equation increasing the value of the frozen coupling. If a
nonzero number of flavors produces any observable effect, this
one should act in the same sense for both solutions. Therefore,
we do not expect large changes in our results with the inclusion
of fermion loops in the SDE solutions, and we can say that the
phenomenological data on $\alpha_s (0)$ is only compatible with
the running coupling determined in Ref.\cite{cornwall}.

It is known that different approximations in the same set of SDE
produce different results. For example, Atkinson and Bloch solved
the same equations of Alkofer {\it et. al} using bare truncation
and performing an angular averaging of the integrals. In this
calculation they obtained $\alpha_s(0)=11.47$ \cite{atk}. When the
angular integrals were performed exactly they found
$\alpha_s(0)\approx 4.2$ \cite{atk-bloch}. In all these cases, the
incompatibility with phenomenological data is still present. These
studies can be improved requiring multiplicative renormalizability
of gluon and ghost propagators \cite{bloch}.

It has been claimed that the asymptotic pion form factor is quite
dependent on the behavior of $\alpha_s$\cite{ji}. Therefore, in
Section III we computed $F_\pi (Q^2)$ with both SDE solutions.
Again, one of the solutions is clearly preferred than the other.
Although we followed a traditional calculation performed by
several authors, where the form factor was calculated using the
nonperturbative running coupling, we commented in Section IV that
a consistent treatment is obtained only if the nonperturbative
gluon propagators are also taken into account. We modified the
expression for the pion form factor including the full gluon
propagator. The pion form factor is clearly modified in the
infrared in both cases compared to the result of the previous
section. It is important to recall that the perturbative QCD
expression for $F_\pi (Q^2)$ is not reliable for small $Q^2$.
However, for large $Q^2$ the incompatibility of one of the
solutions with the data is apparent.

In summary, the phenomenological data on the low energy behavior
of the QCD coupling constant can be used to constrain the
solutions of Schwinger-Dyson equations for the coupling constant
and gluon propagators. More data is necessary, but the ones
already existent indicate that some approximations made in the SDE,
leading to a particular value of the running coupling in the
infrared region, may not be precise enough to reveal its actual behavior.

\section*{Acknowledgments}

This research was supported by the Conselho Nacional de
Desenvolvimento Cient\'{\i}fico e Tecnol\'ogico (CNPq) (AAN), by
Funda\c{c}\~ao de Amparo \`a Pesquisa do Estado de S\~ao Paulo
(FAPESP) (ACA,AAN), by Coordenadoria de Aperfei\c coamento do
Pessoal de Ensino Superior (CAPES) (AM) and by Programa de Apoio
a N\'ucleos de Excel\^encia (PRONEX).

\end{document}